\documentclass[acmtog]{acmartmod}

\usepackage{enumitem}
\usepackage{amssymb}
\usepackage{amsmath}
\usepackage{amsthm}
\usepackage{wrapfig}
\usepackage{overpic}
\usepackage{graphicx}

\newcommand{\bt}{\mathbf{t}}

\newtheorem{prop}{Proposition}

\setitemize[0]{leftmargin=16pt}

\begin{document}

\citestyle{acmauthoryear}
\setcitestyle{square}

\title{On Discrete Conformal Seamless Similarity Maps}

\author{Marcel Campen}
\author{Denis Zorin}
\affiliation{
  \institution{New York University}
}

\begin{abstract}
  An algorithm for the computation of global discrete conformal parametrizations with prescribed global holonomy signatures for triangle meshes was recently described in \cite{Campen:2017:SimilarityMaps}.  In this paper we provide a detailed analysis of convergence and correctness of this algorithm. We generalize and extend ideas of
  \cite{Springborn:2008} to show a connection of the algorithm to Newton's algorithm applied to solving the system of constraints on angles in the parametric domain,
and demonstrate that this system can be obtained as a gradient of a convex energy.
\end{abstract}

\thanks{This work is supported by the National Science Foundation,
  under grant IIS-1320635.}


\setcopyright{none}

\maketitle

\section{Introduction}

In \cite[Sec.\ 7.3]{Campen:2017:SimilarityMaps} we introduce an iterative algorithm that computes a \emph{discrete conformal seamless similarity map} with a prescribed \emph{holonomy signature}, i.e., rotation angles around singularities and along homology loops; a precise definition is given below. 

In this paper we examine the properties of this algorithm in more detail.
We establish the following properties: 
\begin{itemize}
\item If the algorithm converges, it yields a discrete conformal seamless similarity map, and this map has the prescribed signature.
\item A slightly modified version of the algorithm converges unless an infinite sequence of edge flips occurs.
\end{itemize}
The question whether such an infinite sequence can actually occur remains to be answered, similar to the closely related question posed by \cite{Luo:2004} in the context of discrete Ricci flow.  

\section{Background}

\subsubsection*{Discrete surface}

Let $M = (V,E,F)$ be a closed orientable manifold triangle mesh of genus $g$, i.e.\ with $|V|+|F|-|E| = 2-2g$.
Let $M_c = (V_c, E_c, F_c=F)$ be a triangle mesh obtained by cutting $M$ to disk topology using a cut graph $c$ (consisting of edges of $M$).

\subsubsection*{Discrete metric}

Let $G$ be a discrete metric on $M$, i.e.\ an assignment $E \rightarrow \mathbb{R}^{>0}$ of a positive length $l_i^G$ to each edge $i$. This metric carries over to $M_c$ in a trivial manner.

\subsubsection*{Discrete conformal metric}

Let $G'$ be a discrete metric on $M_c$ that is discrete conformally equivalent to $G$ on $M_c$, in the sense that there exists a discrete 0-form $\phi$ on $M_c$ such that 
\begin{equation}  
l_i^{G'} = l_i^G e^{(\phi_v + \phi_w)/2}
\label{eq:elen}
\end{equation}  
for all edges $i\in E_c$, where $i=(v,w)$ is the edge between vertices $v$ and $w$. In the following we simply write $l_i$ instead of $l^{G'}_i$.

\subsubsection*{Discrete conformal map}

Under the assumption that $G'$ respects the triangle inequality for every face of $F$, let $\Theta_r$ denote the sum of inner triangle angles incident at vertex $r$ of $M$ under $G'$.
A vertex $r$ where $\Theta_r = 2 \pi$ is called \emph{regular} under $G'$, otherwise \emph{irregular} (or \emph{extraordinary}).
If all inner vertices of $M_c$ (i.e.\ those not on the boundary due to the cut) are regular, the discrete metric $G'$ is \emph{flat}; it thus implies a (continuous, locally injective) discrete conformal map $f: M_c \rightarrow \mathbb{R}^2$, unique up to a rigid transformation.

\subsubsection*{Discrete seamless similarity map}

A discrete map $f: M_c \rightarrow \mathbb{R}^2$ is called a discrete \emph{seamless similarity} map \protect\cite{Campen:2017:SimilarityMaps} if
there exists a similarity transform $\sigma_i(\mathbf{x}) = s_i R_i\mathbf{x} + \bt_i$, where $R_i$ is a rotation by some integer multiple of $\frac{\pi}{2}$, per cut edge $i$ of $M$ that identifies the two images of $i$ under $f$, and these transforms fulfill a \emph{cycle condition}: around each regular vertex of $M$ the composition of these transforms across incident cut edges (in consistent orientation,
e.g.\ clockwise) is the identity.

\subsubsection*{Dual cycle geodesic curvature}

\begin{figure}[t]
\begin{overpic}[width=0.99\columnwidth]{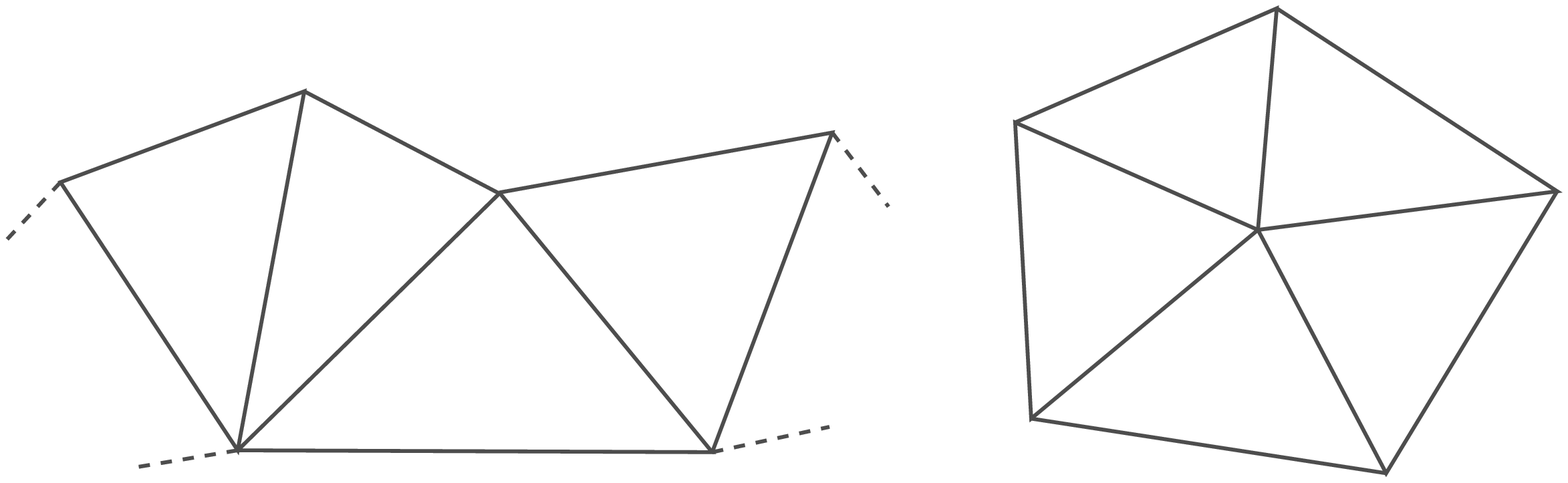}
\put(9.3,9.8){\small$-\alpha^s_3$}
\put(15.1,8.6){\small$-\alpha^s_4$}
\put(28,13){\small$+\alpha^s_5$}
\put(41,7){\small$-\alpha^s_6$}
\put(10.5,14.5){\small$T^s_3$}
\put(19.5,14.5){\small$T^s_4$}
\put(29.5,6.5){\small$T^s_5$}
\put(42.5,12.5){\small$T^s_6$}
\put(83,13){\small$\alpha_a$}
\put(83,18.5){\small$\alpha_b$}
\put(76,19){\small$\alpha_c$}
\put(73,15){\small$\alpha_d$}
\put(78,11){\small$\alpha_e$}
\end{overpic}
\caption{Illustration of angles summed along a dual cycle, i.e.\ triangle strip $T^s$ (left), and around a vertex $r$ (right), where $N(r) = \{a,b,c,d,e\}$.}
\label{fig:alphastrip}    
\end{figure}

Let $\gamma_s$ be a dual cycle of $M$, i.e.\ a directed cyclic triangle strip \cite{Crane:2010}.
Let the $n_s$ triangles forming this cycle be denoted, in sequence, $T^s_m$, $m=1\ldots n_s$. The total geodesic curvature $\kappa_s$ of the cycle $\gamma_s$ under $G'$ can then be expressed as
\begin{equation}  
 \kappa_s = \sum_{m=1...n_s} d^s_m \alpha^s_m,
\label{eq:xxx1}
\end{equation}  
where $\alpha^s_m$ is the angle (under $G'$) of the triangle $T_m^s$ at the vertex that is incident to both, the preceding and succeeding triangle in triangle strip $T^s$, and the sign $d_m = \pm 1$
is determined by whether this vertex is to the left or to the right of the strip, with respect to the direction of the strip (cf.\ Figure \ref{fig:alphastrip} for an illustration).

For the special case of a dual cycle $\gamma_r$ enclosing a single vertex $r$ we can write more succinctly
\begin{equation}  
\kappa_r = \Theta_r = \sum_{i \in N(r)} \alpha_i,
\label{eq:xxx2}
\end{equation}
where $N(r)$ is the set of indices $i$ of inner triangle angles $\alpha_i$ incident at $r$ in $M$ (cf.\ Figure \ref{fig:alphastrip} right), and $\Theta_r$ as above.

\subsubsection*{Dual cycle basis}

Let $\Gamma = \{\gamma_1, \dots, \gamma_{|V|-1+2g}\}$ be a set of dual cycles of $M$ that forms a basis of all dual cycles; concretely, we can choose $|V|-1$ elementary dual cycles around single vertices and $2g$ non-contractible dual cycles \cite{Crane:2010}.

\subsubsection*{Holonomy signature}

The (holonomy) signature of a discrete map $f$ is a collection of $|V|-1+2g$ values $k_s$, defined relative to a basis $\Gamma$ as
$k_s = \kappa_s / \frac{\pi}{2}$.
Note that this signature captures the value $\kappa$ for \emph{any} dual cycle (via linear combination of $\kappa$ values of basis cycles the cycle decomposes to).

An important observation is that, if the map $f$ is a discrete seamless similarity map, $\kappa_s$ is a multiple of $\frac{\pi}{2}$ for any cycle $\gamma_s$ (as well as $\Theta_r$ for any vertex $r$), i.e.\ the signature values $k_s$ are integer numbers.

\subsection{Outline}

Conceptually, the algorithm \cite[Sec.\ 7.3]{Campen:2017:SimilarityMaps} proceeds by constructing a discrete closed 1-form $\xi$ on $M$, such that the implied parametrization satisfies holonomy constraints.  For a suitably chosen cutgraph $c$, on the simply-connected domain $M_c$ this closed 1-form is exact and can be integrated, yielding a discrete 0-form $\phi$ on $M_c$. This 0-form $\phi$ then induces $G'$ via~\eqref{eq:elen}.

We show the following:

\begin{enumerate}

\item if the angles $\alpha_i$ under $G'$ satisfy the set of signature dependent constraints \eqref{eq:discr-constraints},
then $G'$ defines a discrete conformal map (up to rigid transformation) and this map is a seamless similarity map with this given signature,

\item these angles $\alpha_i$ under $G'$ can be computed directly from $\xi$; they are independent of the choice of the cutgraph and the constant of integration,
 
\item the iterative algorithm \cite[Sec.\ 7.3]{Campen:2017:SimilarityMaps} is, in its core, in fact Newton's algorithm applied to solve the non-linear system of equations \eqref{eq:discr-constraints} in the variables $\xi$.

\end{enumerate}

\noindent Hence, \emph{if} the algorithm converges, it yields a discrete 1-form $\xi$, which gives rise to a discrete 0-form $\phi$, that defines a discrete conformally equivalent metric $G'$.
This metric implies a discrete, piecewise linear, seamless similarity map for $M$ that satisfies the holonomy signature constraints. Next, we consider convergence:

\begin{enumerate}

\item[(4)] there is an energy $E$ associated with the mesh $M$, such that
the constraints \eqref{eq:discr-constraints} are the components of this energy's gradient,
for a particular choice of basis for closed 1-forms on~$M$.
This energy is convex.
\end{enumerate}

\noindent The use of a globally convergent variant of Newton's method (with line search, trust region, etc.) \cite[Ch.\ 6]{Newton} would thus allow to guarantee convergence---unless edge flips can cause an issue (cf.\ Section \ref{sec:flips}).

\section{Constraint System}

A discrete metric $G'$ respects a given signature $\{\hat\kappa_s = k_s\frac{\pi}{2}, \hat\Theta_r = k_r\frac{\pi}{2}\}$, if and only if $\kappa_s = \hat\kappa_s$ and $\Theta_r = \hat\Theta_r$, i.e.\ the angles $\alpha_i$ solve this homogeneous system of equations:
\begin{equation}
  \begin{split}
  &\,\,\sum_{i \in N(r)} \alpha_i - \hat{\Theta}_r = 0,\; \mbox{for all vertices $r$}\\
  &\sum_{m=1...n_s} d^s_m \alpha^s_m - \hat{\kappa}_s = 0, \; \mbox{for all cycles $s$}
  \end{split} 
\label{eq:discr-constraints}
\end{equation} 
If the cutgraph $c$ is chosen such that all vertices $r$ with $\hat\Theta_r \neq 2\pi$ (\emph{irregular} vertices) lie on the boundary of $M_c$ (which we can and will always assume in the remainder), this metric $G'$ is flat on $M_c$ and thus implies a map $f$ to the plane. 

\section{Discrete seamless similarity maps from discrete forms}

We can now show that if the discrete metric $G'$, in addition to satisfying \eqref{eq:discr-constraints}, is conformally equivalent to $G$, then this map $f$ is a conformal seamless similarity map.

\begin{prop} Suppose 
for a choice of discrete closed 1-form $\xi$ on $M$, the 0-form $\phi$ on $M_c$ obtained by integration yields a discrete metric $G'$ via \eqref{eq:elen} that respects the triangle inequality for each triangle. If this discrete metric $G'$ satisfies \eqref{eq:discr-constraints}, then it
defines a discrete conformal seamless similarity
map $f$ of $M_c$ with the designated signature.
\end{prop}

\begin{proof}
Let $\sigma_i(\mathbf{x}) = s_i R_i\mathbf{x} + \bt_i$ be a similarity transform per cut edge $i$, uniquely determined by the images of the two copies of this edge: it maps the one image onto the other. We show: around each regular vertex, the composition of these transforms across incident cut edges (in consistent order,
e.g.\ counterclockwise) is the identity.
We use the example vertex depicted in Figure \ref{fig:tricut} with three incident cut edges $a, b, c$, thus transforms $\sigma_a, \sigma_b, \sigma_c$. The common case of two incident cut edges follows directly as a special case, and generalization to even more incident cut edges is trivial.

\begin{figure}[b]
\begin{overpic}[width=0.37\columnwidth]{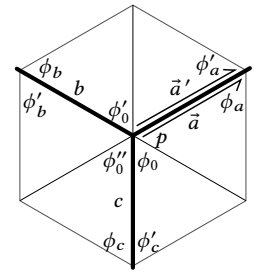}
\put(33,9){\small$\phi_c$}
\put(46,9){\small$\phi_c'$}
\put(33,39){\small$\phi_0''$}
\put(46,39){\small$\phi_0$}
\put(35,58){\small$\phi_0'$}
\put(37,24){\small$c$}
\put(22,66){\small$b$}
\put(65,53){\small$\vec a$}
\put(58,66){\small$\vec a\,'$}
\put(77,61){\small$\phi_a$}
\put(68,75.5){\small$\phi_a'$}
\put(3,61){\small$\phi_b'$}
\put(9,74){\small$\phi_b$}
\put(53,48){\small$p$}
\end{overpic}
\caption{Example cut configuration around a regular vertex. Cuts are bold.}
\label{fig:tricut}    
\end{figure}  

We note that $\phi_a-\phi_0 = \phi_a'-\phi_0'$  ($=\xi_{(0,a)}$, as $\xi$ is defined on $M$), thus $\phi_a'-\phi_a = \phi_0'-\phi_0$. Therefore, by the above definition of edge lengths under $f$, for the scale factor $s_a$ we have $\ln s_a = \phi_0'-\phi_0 = \phi_a'-\phi_a$. Analogously, we get $\ln s_b = \phi_0''-\phi_0'$ and $\ln s_c = \phi_0-\phi_0''$. Hence, $\ln s_a + \ln s_b + \ln s_c = 0$, thus $s_a s_b s_c = 1$.

Let $Df$ be the differential of $f$. Then for the rotations we have $R_a Df(\vec{a}) = Df(\vec{a}')$, $R_b Df(\vec{b}) = Df(\vec{b}')$, $R_c Df(\vec{c}) = Df(\vec{c}')$. Furthermore, let rotations $R_{a'b}, R_{b'c}, R_{c'a}$ be defined via $R_{a'b} Df(\vec{a'}) = Df(\vec{b})$ and so forth. Then $R_{c'a} R_c R_{b'c} R_b R_{a'b} R_a Df(\vec{a}) = Df(\vec{a})$, thus it follows that $ (R_{c'a}R_{b'c} R_{a'b} )(R_c R_b  R_a) = I$.
Due to first set of conditions in \eqref{eq:discr-constraints}, we know that at regular vertices angles under $G'$ sum to $2\pi$, thus $R_{c'a}R_{b'c} R_{a'b} = I$. It follows that $R_c R_b  R_a = I$.

Finally, as $\sigma_c \circ \sigma_b \circ \sigma_a f(p) = f(p)$, and $ \sigma_c \circ \sigma_b \circ \sigma_a f(p) = s_a s_b s_c R_a R_b R_c f(p) + R_a R_b \bt_c + R_a \bt_b + \bt_a = f(p) + R_a R_b \bt_c + R_a \bt_b + \bt_a$, we conclude that $R_a R_b \bt_c + R_a \bt_b + \bt_a = 0$, thus $\sigma_c \circ \sigma_b \circ \sigma_a = I$.

Notice that in the special case of a vertex with two incident cut edges, the transforms on both edges are the same (or, in cyclic orientation, inverses of each other). Thus along a branch of the cut the transition is constant.
That the rotational part of these constant transforms is a multiple of $\pi/2$ follows
from the second set of conditions in \eqref{eq:discr-constraints} when considering a dual cycle $\gamma$ (possibly composed from basis cycles) that crosses the cut: the rotation between the two images of the cut edge is $\hat\kappa_\gamma$.

We conclude that, as the transforms $\sigma$ are similarities that identify the cut images by definition and satisfy the cycle condition, the map $f$ obtained from $\xi$
is a discrete seamless similarity map.

As $G'$ satisfies \eqref{eq:discr-constraints}, this resulting map has the desired signature.
\end{proof}

\section{Angles $\alpha$ from 1-form $\xi$}

The angles $\alpha_i$  non-linearly depend on the edge lengths $l_i$ of $G'$, which in turn depend on $\phi$, which in turn depends, due to an arbitrary choice of cut and arbitrary constant of integration not
even uniquely, on $\xi$. However, we observe that
the angles $\alpha_i$ themselves are uniquely defined by $\xi$ (independent of the cut and the constant of integration).
This allows us to express \eqref{eq:discr-constraints} directly in terms of the 1-form $\xi$.

\paragraph{Notation.}
We consider closed discrete 1-forms $\xi$ on $M$.  We number directed \emph{halfedges} on the mesh, and associate a variable $\xi_i$ with each halfedge.
We use the following notation: for a halfedge $i$, $i'$ is the sibling halfedge, corresponding to the same edge;
$h(i)$ and $t(i)$ are the halfedge head and tail vertices, respectively. 
Each halfedge $i$
\begin{wrapfigure}{r}{0.16\columnwidth}
\vspace{-2mm}\hspace{-5mm}
\begin{overpic}[width=0.21\columnwidth]{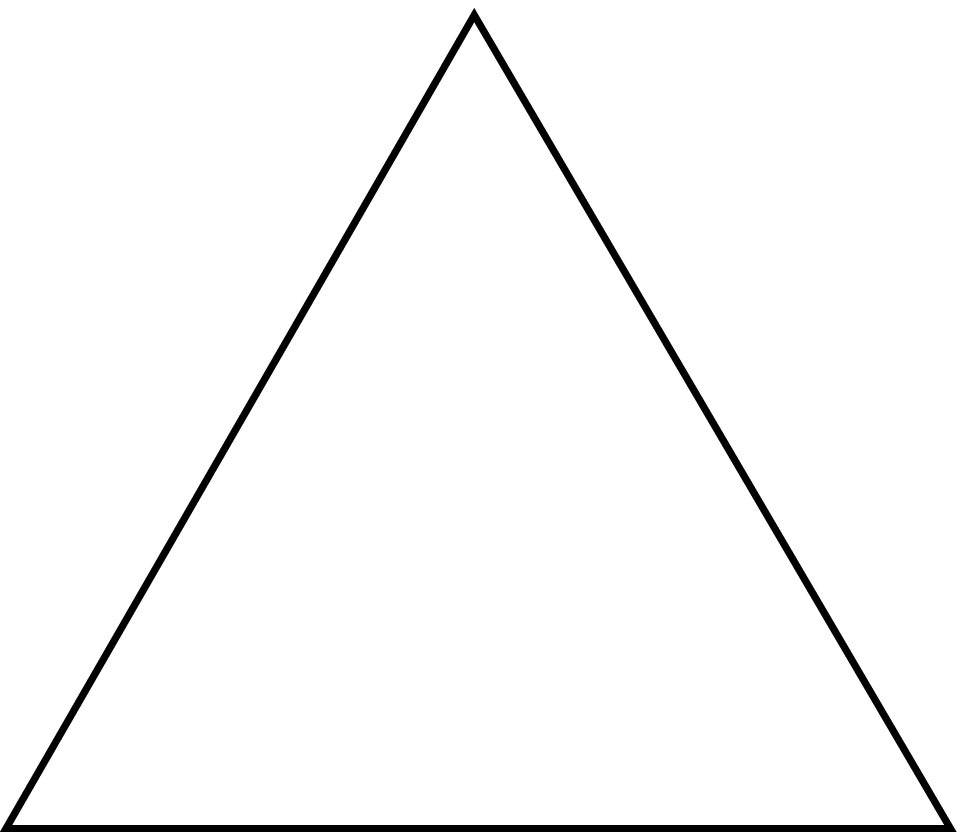}
\put(46,7){$i$}
\put(46,65){$i$}
\put(62,35){$j$}
\put(14,7){$j$}
\put(78,6){$k$}
\put(30,35){$k$}
\end{overpic}
\end{wrapfigure}
belongs to a unique triangle, it is oriented counterclockwise, and the opposite angle in this triangle is denoted~$\alpha_i$.
For a value assignment $\xi$ to define a 1-form, we require
$\xi_{i'} = - \xi_i$. The 1-form is closed if for a triangle with halfedges  $(i,j,k)$ it holds  $\xi_i + \xi_j + \xi_k = 0$.

\begin{prop}
Suppose a discrete 0-form $\phi$
is computed by integration of a closed 1-form $\xi$  on $M_c$, so that $\xi_i = \phi_k-\phi_j$ on
each triangle $(i,j,k)$, and the edge lengths $l_i = l_i^Ge^{(\phi_j + \phi_k)/2}$ satisfy the triangle inequality for every triangle..
Then the angles $\alpha_i$ implied by these edge lengths $l$ depend on the 1-form $\xi$ only,
not on the choice of the cut or the constant of integration.
\end{prop}
\begin{proof}
For $M_c$, the (double) logarithmic edge lengths $\lambda_i$ are defined by $\lambda_i = 2\ln l_i = \lambda^G_i + \phi_j + \phi_k$, with
$\lambda^G_i = 2 \ln l^G_i$.

Furthermore, define the value $\mu_i = \lambda_i^G + \frac{1}{3}(\xi_k-\xi_j)$. We have $\mu_i = 
\lambda_i^G + \phi_j + \phi_k -2s = \lambda_i-2s$, where $s = \frac{1}{3}(\phi_i + \phi_j + \phi_k)$.
In other words, $\mu_i$ (computed from $\xi$) differs from $\lambda_i$ (computed from $\phi$) by $-2s$, a common scaling factor for all three edges of a triangle. As uniformly scaling a triangle does not change its angles, $\alpha_i$ can be computed from the edge lengths $e^{\frac{1}{2}\mu_i},e^{\frac{1}{2}\mu_j},e^{\frac{1}{2}\mu_k}$ (which form a triangle that is \emph{similar} to the triangle formed by $l_i,l_j,l_k$). 
\end{proof}

\section{Newton's method}

Let $L(\xi) = 0$ denote the homogeneous equation system \eqref{eq:discr-constraints}. Note that the left hand side $L$ is a nonlinear function of $\xi$.

Newton's method for finding a solution to $L(\xi) = 0$ makes use of the Jacobian $J_L$ and, starting from the initialization $\xi^0 = 0$, proceeds by iterating:
\begin{equation}
\xi^{n+1} \leftarrow \xi^n -J_L(\xi^n)^{-1} L(\xi^n).
\label{eq:Newton}
\end{equation}
We compare this to the central line of the iterative algorithm given in \cite[Sec.\ 7.3]{Campen:2017:SimilarityMaps}
\begin{equation}
\xi \leftarrow A(G')^{-1}b(G'),
\label{eq:ourNewton}
\end{equation}
(where $\xi^n$ vanishes because it is 'baked' into the metric $G'$ after each step by method \textsf{rescale}, as a merely technical difference for notational simplicity), and show in the following that the matrix $A$ used in the algorithm  equals $J_L$, and the vector $b$ equals $-L$.

\begin{prop}
The iterative algorithm of \cite[Sec.\ 7.3]{Campen:2017:SimilarityMaps}   is (except for the degeneracy handling) Newton's method applied to the nonlinear system \eqref{eq:discr-constraints} in the variables $\xi$.
Thus, as established by the previous propositions, if it converges, the resulting $\xi$ satisfies \eqref{eq:discr-constraints} and, for any choice of cut, defines a discrete conformal seamless similarity map with the given signature.
\end{prop}  

\begin{proof}
Observe, following the derivation of \cite[Eq.\ 14]{Springborn:2008}, that for a triangle $(i,j,k)$ the first-order change in $\alpha_i$ relative to $\xi$ is $$d \alpha_i = \frac{1}{2}\cot\alpha_k d\xi_k - \frac{1}{2} \cot\alpha_j d\xi_j $$
Thus, the linearized form of  the first type of constraint in  \eqref{eq:discr-constraints} can be written as
\begin{equation}
\sum_{i \in N(r)} \frac{1}{2} \cot \alpha_k \Delta \xi_k - \frac{1}{2} \cot \alpha_j \Delta \xi_j = \hat{\Theta}_r-\Theta_r,\; \mbox{for every $r$},
\label{eq:linv}
\end{equation}
where $j,k$ are the other two halfedge indices in the triangle containing halfedge $i$.
Note that $\Delta \xi$ here denotes the Newton step, the change in the 1-form $\xi$ that is solved for in each iteration of \eqref{eq:Newton} (called $\xi$ in \eqref{eq:ourNewton}, because $\xi^n=0$ in that algorithm).
Let $O(r)$ be the set of outgoing halfedges at vertex $r$, i.e.\ halfedges $i$ with $t(i) = r$. With this and $\xi_i = -\xi_{i'}$, \eqref{eq:linv} can be rearranged to
\begin{equation}
\sum_{i \in O(r)} \frac{1}{2}(\cot \alpha_i + \cot \alpha_{i'}) \Delta \xi_i = \hat{\Theta}_r-\Theta_r,\; \mbox{for every $r$}.
\tag{\theequation${}^\prime$}
\label{eq:linv2}
\end{equation}

Similarly, the second type of constraints in  \eqref{eq:discr-constraints} is linearized as
\begin{equation}   
\sum_{m=1...n_s} \!\!\! d^s_m \frac{1}{2}(\cot \alpha_k \Delta \xi_k - \cot \alpha_j \Delta \xi_j) = \hat{\kappa}_s - \kappa_s, \; \mbox{for every $s$},
\label{eq:linl}
\end{equation}
where $j,k$ are the two halfedge indices in triangle $T_m^s$ not opposite to angle $\alpha_m^s$.
Let $E(s)$ denote the set of all halfedges between successive triangles $T^s_m$, $T^s_{m+1 \text{ mod } n_s}$ pointing consistently from the left boundary of the strip to the right boundary (in accordance with the definition of the sign $d_m^s$). With this, and grouping factors per halfedge in $E(s)$, we can rearrange \eqref{eq:linl} to
\begin{equation}
\sum_{i \in E(s)} \frac{1}{2}(\cot \alpha_i + \cot \alpha_{i'}) \Delta \xi_i = \hat{\kappa}_s - \kappa_s,\; \mbox{for every $v$}.
\label{eq:linl2}
\tag{\theequation${}^\prime$}
\end{equation}

Realizing that, in terms of \cite{Campen:2017:SimilarityMaps}, $\kappa_s \!=\! \kappa^{tot}[\gamma^d_s]$ and $\Theta_r =  2\pi - K_{G}[r]$, one observes that \eqref{eq:linv2} and \eqref{eq:linl2} equal equation system 6 in \cite{Campen:2017:SimilarityMaps}, which defines $A$ and $b$. Thus \eqref{eq:ourNewton} is indeed a Newton step for solving \eqref{eq:discr-constraints}.
\end{proof}  

\section{Underlying convex energy}
\label{sec:app:convex}

Analogously to the simply-connected case \cite{Springborn:2008} (where only elementary dual cycles, i.e.\ constraints of the first type in \eqref{eq:discr-constraints} play a role), $L$, the vector of constraint left-hand sides, can be shown to be a gradient of a
convex function $E$. This suggests that global convergence of the algorithm can be ensured if one augments it with a line search (or other global convergence techniques, such as a trust region approach  \cite[Ch.\ 6]{Newton})---assuming triangle inequalities do not get violated (cf.\ Section \ref{sec:flips}).

To show this, we introduce an auxiliary variable $\psi_i$ per halfedge, satisfying $\psi_i = \frac{1}{3}(\xi_k - \xi_j)$ in the triangle $(i,j,k)$.
Note that then $\mu_i = \lambda^G_i +  \psi_i$. We  also observe that, due to closedness, $\xi_i =(\psi_l - \psi_k)$.

We define a triangle function $g(\psi_i, \psi_j, \psi_k)$, similar to the function $f$ defined in \cite[Eq.\ 8]{Springborn:2008} involving Milnor's Lobachevsky function 
$\Lambda$,
 but depending on the values of $\psi$ (and thus, by a linear change of variable, $\xi$) only.
\begin{equation}
g(\psi_i, \psi_j, \psi_k) = \sum_{\ell\in {i,j,k}} (\lambda_\ell^G + \psi_\ell)\alpha_\ell + 
\Lambda
(\alpha_\ell),
\label{eq:magic-function}
\end{equation}
It differs from the function $f$ by a linear change of variables and addition of a linear function, thus, it is convex as well.
This function has the following property, similar to the corresponding property of $f$:
\begin{equation}
\partial_{\psi_i} g(\psi_i, \psi_j, \psi_k)  = \frac{1}{2}\alpha_i
\label{eq:magic-function-deriv}
\end{equation}
with similar equations for derivatives with respect to $\psi_j$ and $\psi_k$ obtained by cyclic permutation of $(i,j,k)$.
This property makes it possible to formulate conditions on angles (such as in \eqref{eq:discr-constraints}) as components
of the gradient of an energy constructed from $g(\xi_i, \xi_j, \xi_k)$ for individual triangles. Towards this goal we make the following general observation.  

\subsubsection*{General observation}
Suppose for an energy $E = E(x)$, with the variables $x$ satisfying constraints $Cx = 0$, we make a  change of variables $x = Py$, where the variables $y$ are independent, i.e.,  the columns of
$P$ are a basis of the null space of $C$.  Then $\nabla_y E(Py) = P^T \nabla_x E(x).$  Let $P^+ = (P^T P)^{-1} P^T$ be the pseudoinverse of $P$.
Then $y = P^+ x$. 

Consider an augmented energy $E'(x) = E(x) - b^T P^+ x = E(Py) - b^T y$.  Then $\nabla_y E' = P^T \nabla_x E - b$.  Let $p_i$ be the columns of $P$ (rows of $P^T$). We conclude that for any choice of basis $y$ in the null-space of the constraint
matrix $C$, the value of $x^*$ corresponding to an extremum of $E(x)$, satisfies the constraints $p_i \nabla_x E(x^*) - b_i = 0$.

We apply this observation in our setting to show how to construct an energy function that yields the vector of left-hand sides of  \eqref{eq:discr-constraints} as a gradient.

\begin{figure}[t]
\begin{overpic}[width=0.99\columnwidth]{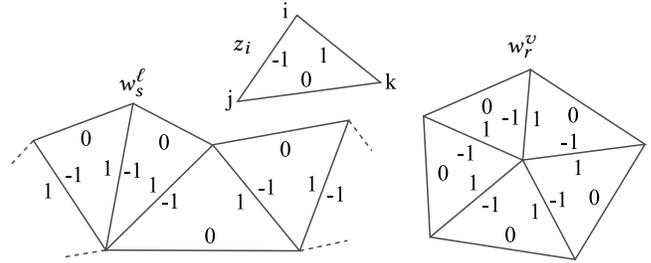}
\put(35,33.3){$z_i$}
\put(78,33){$w_r^v$}
\put(17,27){$w_s^\ell$}
\end{overpic}
\caption{Illustration of employed basis elements for closed 1-forms, $w_s^\ell$ and $w_r^v$, as well as basis elements $z_i$ for $\psi$.}
\label{fig:stripbasis}    
\end{figure}

\paragraph{Basis change}
We define a basis for closed 1-forms  $w^v_r$, $w^\ell_s$, $r=1\ldots |V|-1$, $s = 1\ldots 2g$, jointly denoted by $w_t$, $t = 1\ldots |V|+2g-1$.
$w^v_r$ is obtained by applying the discrete exterior derivative to the hat function centered at a vertex $r$,
i.e. each halfedge $i$ with $h(i) = r$ is assigned value $1$, and $t(i) = r$ is assigned $-1$ (cf.\ Figure \ref{fig:stripbasis}).

For a dual cycle $\gamma_s^d$, $w^\ell_s$ is defined by setting each interior halfedge of the cycle to $1$
if it crosses the midline from right to left (relative to the direction of the cycle), and $-1$ otherwise, forming vector $w^\ell_s$.
Together, the vertex and cycle basis forms are denoted $w_t$, $t=1\ldots |V|-1+2g$.

For a closed 1-form $\xi$, we denote the coefficients of the form represented in this basis by $y_t$,  forming a vector $y$, $\xi = Wy$, where $W$ is the matrix with columns $w^t$.

We also consider the basis $z_i$, $i=1\ldots 3|F|$, with basis forms associated with vertices,  corresponding to the variables $\psi_i$, which have $-1$ on the halfedge $j$, 1 on the halfedge $k$ and 0 elsewhere. Then $\xi = Z\psi$, where $Z$ is the matrix with columns $z_i$. 

We observe that $w^v_r = \sum_{i\in N(r)} z_i$, and $w^\ell_s =  \sum_{m=1^m_\ell} d^\ell_m z^\ell_m$, with $d^\ell_m$ defined as above, and $z^\ell_m$ are the basis functions corresponding to vertices on the boundary of the cycle triangle strip $T_m^\ell$. Notice that with these basis choices, the coefficients for expressing $w_t$ in terms of $z_i$ are exactly the coefficients in \eqref{eq:discr-constraints}, expressing total angles in terms of angles $\alpha_i$.

In a general form, we write these relations as $w_t = Z p_t$, i.e.,  $W = Z P$, where the columns of $P$ are $p_i$.  From this we infer that $\xi = Wy  = ZPy$; comparing to $\xi = Z\psi$, we conclude that $\psi= Py$.

These definitions allow us to formulate the following proposition:
\begin{prop}
Define the energy
$$E(\psi) = \sum_{T_{ijk}}  g(\psi_i, \psi_j,\psi_k),$$
and let $E'(\psi) = E(\psi)- \frac{1}{2}b^T P^+ \psi$,
where the components of $b_t$ of $b$ are, for $t = 1\ldots |V|-1$, the target angle sums $\hat{\Theta}_t$ at vertices, excluding one (implied by the theorem of Gauss-Bonnet), 
and for $t = |V|\ldots |V|+2g-1$, the target geodesic curvatures $\hat{\kappa}_t$ on dual cycles $\gamma_{t-|V|+1}$. The matrix $P^+$ is
the pseudoinverse of $P$ formed by the coefficient vectors $p_t$ of the closed 1-form basis $y_t$, $t=1\ldots |V|-1+2g$, expressed in terms of the basis $z_i$
defined above.  Then the left-hand-sides of \eqref{eq:discr-constraints} are the components of $\nabla_\psi E$.
\end{prop}  

\begin{proof}
Using the relation $\psi = Py$, we rewrite our energy as $E'(\psi)  = E(Py) - \frac{1}{2}b^Ty$, and computing the gradient yields $P^T \nabla_\psi E - \frac{1}{2}b$.
As the columns $p_t$ of $P$ coincide with the coefficients of \eqref{eq:discr-constraints}, and
$\partial_{\psi_i} E = \partial_{\psi_i} g^{T(i)} = \frac{1}{2}\alpha_i$,   where $g^T = g(\psi_i, \psi_j, \psi_k)$ for the triangle $T(i)$ containing the halfedge $i$, each row has the form $\frac{1}{2} p_t \cdot \alpha - \frac{1}{2} b_t$, $t=1\ldots V-1+2g$, which is exactly the left-hand sides of \eqref{eq:discr-constraints}.   
\end{proof}  

\section{Triangle Inequality}
\label{sec:flips}

In our analysis we have generally assumed that the discrete metric $G'$ (at intermediate steps as well as in the end) respects the triangle inequality for every triangle of $M_c$; otherwise angles $\alpha$ are not well-defined.

\cite{Springborn:2008} extend definitions of gradient and Hessian to violating states, thereby allowing their algorithm to proceed even if the triangle inequality is violated at an intermediate step.
If, however, the triangle inequality is violated in the end, the metric $G'$ does not well-define a map. Note that, depending on the mesh connectivity and the desired signature, such violations can be inherent rather than be artifacts of the algorithm; no discrete map with the desired properties might exist on a given mesh.

\subsection*{Preventing violations}
In \cite{Campen:2017:SimilarityMaps} a strategy is used that prevents violations, instead of extending the energy definition to cover them, and that modifies the mesh connectivity where necessary. In a way similar to a line search strategy, the Newton step is truncated if it would lead to a violation, such that instead a degenerate configuration is obtained (that necessarily occurs before a violation occurs), which is immediately resolved by an edge flip. This closely follows a continuous technique proposed by \cite{Luo:2004} to deal with singularities of the discrete Yamabe flow.

\subsection*{Effect on convergence}
These edge flips are performed intrinsically, i.e.\ values $\Theta_r$ and $\kappa_s$, and thus the energy $E$, are preserved.
If finitely many edge flips occur in the course of the algorithm, it obviously proceeds with normal Newton steps after the last one, converging if a globally convergent variant is employed.

Questions left to be answered are:
\begin{itemize}
\item can an infinite sequence of edge flips occur? (note, e.g., that one cannot guarantee meeting Wolfe's curvature condition if occurring degeneracies can truncate the steps in an arbitrary manner). The same question applies to discrete Yamabe flow with edge flips, as discussed in \cite{Luo:2004}.
\item can two adjacent triangles degenerate simultaneously in such a manner that the effected intrinsic edge flip leads to an edge of zero length?
\end{itemize}

\bibliographystyle{ACM-Reference-Format}
\bibliography{DCSSM}


\begin{thebibliography}{00}


\ifx \showCODEN    \undefined \def \showCODEN     #1{\unskip}     \fi
\ifx \showDOI      \undefined \def \showDOI       #1{{\tt DOI:}\penalty0{#1}\ }
  \fi
\ifx \showISBNx    \undefined \def \showISBNx     #1{\unskip}     \fi
\ifx \showISBNxiii \undefined \def \showISBNxiii  #1{\unskip}     \fi
\ifx \showISSN     \undefined \def \showISSN      #1{\unskip}     \fi
\ifx \showLCCN     \undefined \def \showLCCN      #1{\unskip}     \fi
\ifx \shownote     \undefined \def \shownote      #1{#1}          \fi
\ifx \showarticletitle \undefined \def \showarticletitle #1{#1}   \fi
\ifx \showURL      \undefined \def \showURL       #1{#1}          \fi
\providecommand\bibfield[2]{#2}
\providecommand\bibinfo[2]{#2}
\providecommand\natexlab[1]{#1}
\providecommand\showeprint[2][]{arXiv:#2}

\bibitem[\protect\citeauthoryear{Campen and Zorin}{Campen and Zorin}{2017}]%
        {Campen:2017:SimilarityMaps}
\bibfield{author}{\bibinfo{person}{Marcel Campen} {and} \bibinfo{person}{Denis
  Zorin}.} \bibinfo{year}{2017}\natexlab{}.
\newblock \showarticletitle{Similarity Maps and Field-Guided T-Splines: a
  Perfect Couple}.
\newblock \bibinfo{journal}{{\em ACM Trans. Graph.\/}} \bibinfo{volume}{36},
  \bibinfo{number}{4} (\bibinfo{year}{2017}).
\newblock


\bibitem[\protect\citeauthoryear{Crane, Desbrun, and Schr{\"o}der}{Crane
  et~al\mbox{.}}{2010}]%
        {Crane:2010}
\bibfield{author}{\bibinfo{person}{Keenan Crane}, \bibinfo{person}{Mathieu
  Desbrun}, {and} \bibinfo{person}{Peter Schr{\"o}der}.}
  \bibinfo{year}{2010}\natexlab{}.
\newblock \showarticletitle{Trivial Connections on Discrete Surfaces}.
\newblock \bibinfo{journal}{{\em Comp. Graph. Forum\/}} \bibinfo{volume}{29},
  \bibinfo{number}{5} (\bibinfo{year}{2010}), \bibinfo{pages}{1525--1533}.
\newblock


\bibitem[\protect\citeauthoryear{Dennis and Schnabel}{Dennis and
  Schnabel}{1996}]%
        {Newton}
\bibfield{author}{\bibinfo{person}{J.~E. Dennis} {and} \bibinfo{person}{Robert
  Schnabel}.} \bibinfo{year}{1996}\natexlab{}.
\newblock \bibinfo{booktitle}{{\em Numerical Methods for Unconstrained
  Optimization and Nonlinear Equations}}.
\newblock \bibinfo{publisher}{Society for Industrial and Applied Mathematics}.
\newblock


\bibitem[\protect\citeauthoryear{Luo}{Luo}{2004}]%
        {Luo:2004}
\bibfield{author}{\bibinfo{person}{Feng Luo}.} \bibinfo{year}{2004}\natexlab{}.
\newblock \showarticletitle{Combinatorial Yamabe flow on surfaces}.
\newblock \bibinfo{journal}{{\em Communications in Contemporary Mathematics\/}}
  \bibinfo{volume}{6}, \bibinfo{number}{05} (\bibinfo{year}{2004}),
  \bibinfo{pages}{765--780}.
\newblock


\bibitem[\protect\citeauthoryear{Springborn, Schr{\"o}der, and
  Pinkall}{Springborn et~al\mbox{.}}{2008}]%
        {Springborn:2008}
\bibfield{author}{\bibinfo{person}{Boris Springborn}, \bibinfo{person}{Peter
  Schr{\"o}der}, {and} \bibinfo{person}{Ulrich Pinkall}.}
  \bibinfo{year}{2008}\natexlab{}.
\newblock \showarticletitle{Conformal equivalence of triangle meshes}.
\newblock \bibinfo{journal}{{\em ACM Trans. Graph.\/}} \bibinfo{volume}{27},
  \bibinfo{number}{3} (\bibinfo{year}{2008}).
\newblock


\end{thebibliography}

 \end{document}